\documentstyle[amssymb,prl,aps,epsf,epsfig,floats,twocolumn]{revtex}
\begin{document}
\draft

\twocolumn[\hsize\textwidth\columnwidth\hsize\csname
@twocolumnfalse\endcsname

\title{
Nucleon-Nucleon Scattering  Observables in Large-$N_c$ QCD}
\author{Thomas D.~Cohen and Boris A.~Gelman} 
\address{Department of Physics, University of Maryland, 
College Park, MD 20742-4111} 
\maketitle 
\begin{abstract} 
Nucleon-nucleon scattering observables are considered in the context of the
large $N_c$ limit of QCD for initial states with moderately high momenta 
($p \sim N_c$). The scattering is studied in the framework of the  
time-dependent mean-field approximation. We focus on the dependence 
of those observables on the spin and isospin of the initial state which may be 
computed using time-dependent mean-field theory. We show that, up to  
corrections, all such observables must be invariant under simultaneous spin 
and isospin flips ({\it i.e.} rotations through $\pi/2$ in both spin and  
isospin) acting on either particle. All observables of this class obtained
from spin unpolarized measurements must be isospin independent up to $1/N_c$
corrections. Moreover, it can be shown that the leading correction is of
relative order $1/N_c^2$ rather than $1/N_c$. 
 
\end{abstract} 
\medskip]

The strong interaction between two nucleons is the basic ingredient of nuclear 
physics. We wish to explore qualitative features of the nucleon-nucleon  
interaction  which may be understood from QCD.  In this context, it may be 
useful to consider the interaction in the limit when the number of colors,
$N_c$, of QCD becomes large \cite{LN1,LN2}, and to treat $1/N_c$ as an
expansion parameter. Certain aspects of QCD can be deduced in this limit in a
model-independent way. For example, the spin-flavor structure of certain  
amplitudes in the single baryon sector may be fixed \cite{SF}. The  
large $N_c$ limit can also be used to determine the leading spin-isospin  
dependence of certain nucleon-nucleon scattering observables.  
 
Baryons in large-$N_c$ QCD were first discussed by Witten, \cite{LN2}, who
argued that  they are well approximated by a Hartree-type mean-field
treatment. This picture works cleanly when dynamical gluons are integrated
out. For the case of light quarks, explicitly deducing the Hartree equations
of motion  is not a tractable problem with present techniques. However,
the $N_c$ dependence of various baryon properties can still be deduced. Thus,
for example, the baryon mass is of order $N_c$ and the baryon size is of order 
unity \cite{LN2}. 
  
Witten also considered the baryon-baryon interaction. He argued that the 
strength of this interaction is of order $N_c$ and the proper framework is the
time-dependent mean-field theory (TDMFT). The argument is similar to the 
single baryon case: each quark moves in an average potential due to the other 
$2\, N_c -1$ quarks which in this context is time-dependent.   
The mean-field nature of the  nucleon-nucleon scattering in the large $N_c$ 
limit imposes constraints on spin-isospin dependence of certain observables. 
One of the limitations of such an approach is that only certain observables
can be calculated in TDMFT. This limitation can be quite severe. For example,
both the elastic and total scattering cross sections are not calculable in
TDMFT.  
 
Witten pointed out that  baryon-baryon scattering observables are expected to
possess a smooth limit only for momenta of order $N_c$. For this reason we
will focus on this regime.  The regime of momenta of order $N_c^0$ has been
considered elsewhere where the stress has been on the nucleon-nucleon
potential \cite{NN}.  
   
Although we do not know the TDMFT equations for large $N_c$ QCD, we know its
general form  and can deduce from this the leading order spin-isospin
dependence of certain obervables. As this form may not be immediately
familiar, it is useful to also derive results in the context of TDMFT for the
Skyrme model. The Skyrme model describes nucleons as topological solitons of a
chiral field $U$ and is believed to capture all of the model-independent
large $N_c$ results of QCD. It has the virtue that the mean-field theory is
trivially equal to the classical field theory. In both the Skyrme model and
the Hartree treatment the mean-field solution for a single  
baryon is a ``hedgehog'' form which breaks both rotational and isorotational  
invariance.  At the mean-field level this means that there exist manifolds of  
degenerate baryon states corresponding to rotated hedgehogs. These manifolds  
are parameterized by the three independent variables which specify a rotation. 
   
One can specify these as being the three Euler angles; an alternative  
parameterization  is to  specify the rotation in terms of a matrix $A$ given
by $A \, \equiv \, a_0 \, + \, \vec{a}\cdot \vec{\tau} \, \in \, SU(2) $,
where $a_0$, $a_1$, $a_2$ and $a_3$ are real collective variables satisfying a 
constraint $a_0^2 + \vec{a} \cdot \vec{a} = 1$. The standard interpretation of
these states at the full quantum level is that the mean-field states
correspond to linear combinations of {\it nearly} degenerate
states---{\it i.e.} states whose masses are split by ${\cal O}(1/N_c)$. The
quantized states can be viewed as collective wave functions 
associated with the variable parameterizing the rotations of the  
hedgehog; up to normalization constants, these collective wave functions  
are simply the Wigner $D$ matrices \cite{Dmat}: 
$\Psi (A) = (J+1/2)^{1/2} \, \pi^{-1}\, D^{I=J}_{m, m_I} (A)$,
where $J$ and $I$ are spin and isospin of the quantized state. 
 
In TDMFT treatments of baryon-baryon scattering,  
the initial conditions are two well-separated baryons moving towards each 
other along some axis, $\hat{n}$ (which in an experimental situation is the 
beam direction), off-set by an impact parameter $\vec{b}$.  
The initial baryon states are (rotated) hedgehogs. The initial 
states can be parameterized by two sets of rotation parameters---one 
associated with the orientation of each initial hedgehog. When a 
quantity, ${\cal O}$, is calculated in TDMFT its value will depend on the 
relative orientation of the two initial hedgehogs and $\vec{b}$. Up to 
$1/N_c$ corrections quantities which are amenable to calculation in TDMFT can 
be written as ${\cal O}(A_1, A_2, \vec{b},\hat{n},E)$, where $A_1$ and $A_2$ 
represent the initial orientations of the two hedgehog baryons, $\hat{n}$ is 
the beam direction, 
$E$ is the initial kinetic energy of each baryon (in the center-of-mass frame) 
and $\vec{b}$ is the impact parameter (the projection perpendicular to 
$\hat{n}$ of the initial displacement of the two particles from each other). 
The orientations, $A_1$ and $A_2$, are associated with particular linear 
combinations of spin and isospin components of the baryons states. By varying 
these orientations one can extract the part of the result which is associated 
with the various spin-isospin configurations of the initial nucleon states.
This allows us to deduce our principal result: all observables which can be 
deduced from TDMFT must be invariant under simultaneous 
spin and isospin flips ({\it i.e.} rotations through $\pi/2$ in both spin and 
isospin) acting on either particle up to $1/N_c$ corrections. 
 
To compute an obseravble from TDMFT requires that at the QCD level the
observable can be expressed directly in terms of an expectation value of  
some Heisenberg-picture operator (which we will denote as ${\cal O}$) in an  
initial plane wave state of two nucleons. All sensitivity to the future time  
evolution is contained in the operator. First one argues 
that in the $p \sim N_c$ regime the expectation value in these plane-wave 
states can be expressed in terms of integrals of expectation values of initial 
states consisting of well localized wave packets of two (rotated) hedgehog 
baryons heading towards each other and off-set by an impact parameter 
$\vec{b}$. The integrations are over $\vec{b}$ and the parameters specifying
the initial rotations of the hedgehogs.  
 
The integrations over $\vec{b}$ is quite standard. For $p \sim N_c$, the
characteristic momentum is much larger than the inverse size of the region of
interaction. Only a small part of the interaction region is probed
coherently \cite{Landau}. The variable $\vec{b}$ is essentially classical and
the various types of cross sections we will use as observables can be written
as an integral over the impact parameter which specifies the section of the
interaction region being probed.   
 
The integration over the hedgehog orientations follows the strategy of
Ref.~\cite{CohBron}. The mean-field theory, which is essentially classical in
nature, accurately describes all of the variables except the collective
ones---{\it i.e.} those degrees of freedom which at the classical level can be
excited with no energetic cost. These variables must be quantized in order to
project onto physical states with the correct quantum numbers. The variables
associated with the collective isorotations of the widely-separated hedgehogs
are collective variables. The way the quantization is realized is simple: one
expresses the observable at the mean-field level in terms of the $A$'s and
their time derivatives $\dot{A}$. The $A_{1,2}$ and $\dot{A} _{1,2}$ are
dynamical variables which can be quantized. 
 
To proceed further, one must analyze the time scale of the interaction. The
interaction takes place in a time of order $N_c^0$:  the size of the baryons
is order $N_c^0$ and in this regime the velocity is also $N_c^0$. Thus it
takes a time of order $N_c^0$ for the two baryons to substantially overlap 
and interact. This in turn, implies that the observable as a function 
$A_{1,2}$ and $\dot{A}_{1,2}$ should be independent of $N_c$. Once the
observable is written as a function of  $A_{1,2}$ and $\dot{A}_{1,2}$, the
next step is to make a Legendre transformation to express the observable as a
function of the $A$'s and their conjugate momenta. The conjugate momenta can
be expressed in terms of the isospin and the intrinsic isospin in the
body-fixed frame which in hedgehog models is the spin. However, in the 
process of making this Legendre transformation, any variables $\dot{A}_{1,2}$ 
get mapped into $\vec{I}_{1,2}/{\cal I}$ and $\vec{S}_{1,2}/{\cal I}$.   
Note, that ${\cal I}$, the moment of inertia, is ${\cal O}(N_c)$ as can be
seen in explicit model calculations, \cite{SK}, and from the model independent
analysis \cite{Jsquare}. Since in our initial states the isospin and spin are
of order $N_c^0$, it is clear that if one does a $1/N_c$ expansion of the
Legendre transformed operator, the leading order terms can depend on the $A$'s
but not on the spins or isospins. Thus at leading order in the $1/N_c$
expansion we can neglect the spin and isospins and treat the operator as a
diagonal operator in $A_{1,2}$. A simple way to understand this is that
because of the large moment of inertia the characteristic rotational period of
the hedgehogs is $N_c$ and, in the large $N_c$ limit, is irrelevant during
reactions taking place during time scales of order unity.      
 
The operators obtained from TDMFT are of the form ${\cal O}(A_1, A_2, \vec{b},
\hat{n},E)$. The variables $A_{1,2}$, $E$ and $\vec{b}$ are fixed by initial
conditions in the TDMFT; $\vec{b}$ will be integrated over in the standard  
manner. The observables which we focus are expectation values in states with 
fixed initial nucleon spins and isospins.  In experiments this is obtained by  
creating the given initial state multiple times and averaging over 
measurements. To compute nucleon matrix elements of  operators which are
functions of $A_{1,2}$, one integrates the operator weighted by the
 appropriate collective wave functions $\Psi(A)$: the expectation value of an 
operator $f(A_1,A_2)$ in a two-nucleon state with spin projections along  
an arbitrary axis $m^{(1,2)}$ and isospin projections ${m_I}^{(1,2)}$ is 
\begin{eqnarray} 
&{}&\langle f 
\rangle_{m^{(1)}, {m_I}^{(1)}; m^{(2)} , {m_I}^{(2)} } \, = \label{exp}\\ 
&{}& \int  {\rm d}A_1 {\rm d}A_2  
\left | D^{1/2}_{m^{(1)},{m_I}^{(1)}}(A_1) \right |^2  
\left | D^{1/2}_{m^{(2)},{m_I}^{(2)}}(A_2) \right |^2 f(A_1,A_2) \,  
\nonumber 
\end{eqnarray} 
where $\int dA$ is the $SU(2)$ invariant measure normalized so that 
$\int dA = 2\,\pi^2$. 
 
The observable is a cross section associated with the ${\cal O}$ for nucleons
with fixed  initial spins and isospins,  
$\sigma^{\cal O }_{m^{(1)} {m_I}^{(1)} m^{(2)} {m_I}^{(2)}}$.  It is 
 obtained by an integration of $\langle {\cal O}( \vec{b},  \hat{n},E) )  
\rangle_{m^{(1)} {m_I}^{(1)} m^{(2)} {m_I}^{(2)}}$  over $\vec{b}$: 
\begin{eqnarray} 
&{}&\sigma^{\cal O}_{m^{(1)} {m_I}^{(1)} m^{(2)} {m_I}^{(2)}} \, = \, 
\label{sig} \\ 
&{}&\, \int \, {\rm d^2} \, \vec{b} \, \langle {\cal O}( \vec{b}, \hat{n},E) ) 
\rangle_{m^{(1)}, {m_I}^{(1)}; m^{(2)} , {m_I}^{(2)} } \nonumber 
\end{eqnarray} 
where corrections are higher order in $1/N_c$. 
 
The central result of our paper is contained in Eqs.~(\ref{exp}) and  
(\ref{sig}). The dependence on the initial spin and isospin projections of the
individual nucleons of the cross section comes about entirely from the Wigner
$D$ matrices inside the integrals of  Eq.~(\ref{exp}). Note, however, that
what is relevant is the square modulus of the $D$ matrices.  There is a
well-known property of the Wigner $D$ matrices 
$\left ( D^J_{m, n} \right)^* = (-1)^{m-n} D^J_{-m ,-n}$ which implies  
$\left |D^{1/2}_{m , n}(A) \right |^2 =  \left |D^{1/2}_{-m , -n}(A) \right  
|^2$. This means that up to $1/N_c$ corrections the observed cross section
will be unchanged if one simultaneously flips both the spin and the isospin of
either particle 1 or particle 2. This strongly constrains the spin and isospin
dependence of the cross section. 
 
To illustrate this constraint,  consider some observable which averages over 
the direction of all detected particles. Thus, apart from the spins, the only
vector left in the problem is the beam axis $\hat{n}$. Suppose further that
the observable in question is time reversal, parity even and isoscalar.
Then from general invariance properties one has 
\begin{eqnarray} 
&\sigma^{\cal O}& \,  =   \, 
A_0 \, + \, B_0 \, (\vec{\sigma}^{(1)} \cdot \vec{\sigma}^{(2)}) \, 
 + \, C_0 \,(\hat{n} \cdot \vec{\sigma}^{(1)})\,  
(\hat{n} \cdot \vec{\sigma}^{(2)})   
\nonumber \\ 
& & + \left ( A_I \, +  
\, B_I \, (\vec{\sigma}^{(1)} \cdot \vec{\sigma}^{(2)}) \, +  
\, C_I \, (\hat{n} \cdot \vec{\sigma}^{(1)})\,  
(\hat{n} \cdot \vec{\sigma}^{(2)}) \right ) 
\nonumber \\ 
& & \times \, (\tau^{(1)}\cdot \tau^{(2)})  
\label{obs1} 
\end{eqnarray} 
where the $A$, $B$ and $C$ coefficients are functions of energy. The leading
order amplitude must be invariant under a simultaneous spin and isospin flip
of either of the two particles. This implies that $A_I = 0$, $B_0=0$,
$C_0=0$ at leading order, since the terms they multiply change signs under
this transformation.  
For a more general observable the same scheme may be employed: one finds  
all vectors in the problem and constructs the most general form the amplitude 
can take contracting the initial spins into all vectors in all ways 
consistent with the symmetry. One then imposes the large $N_c$ rule to 
eliminate all terms which change signs under a simultaneous spin and 
isospin flip of either particle. One immediate consequence of this analysis is 
that all observables in this regime which are obtained from unpolarized 
measurements must be isoscalars.  
 
There are several fundamental limitations in the use of TDMFT to directly
compute scattering observables \cite{TDMFTF2}. TDMFT is a treatment  
of average quantities rather than any particular scattering channel. Thus,  
one cannot directly compute S-matrix elements since S-matrix elements are 
defined for particular  channels. Moreover, the form of  
the initial and final states are constrained by the form of the trial wave  
function which are not rich enough to include the correlations necessary  
to get the translationally invariant time-independent plane-wave asymptotic  
states of the full quantum theory. Thus the initial states are time-dependent
solutions of two lumps moving towards each other while the final states are  
even more complicated.  Clearly, since TDMFT cannot compute the S-matrix 
it can only give sensible information about quantities which average over many
S-matrix elements. Such observables might, for example, be associated with
various kinds of collective flow. Moreover, such flow must be expressible in
terms of operators which do not encode correlations beyond those built into
the mean-field trial states; one can follow the flows of various quantum
numbers but not of particular correlations of individual hadrons.   
 
It is not hard to determine which experimental observables are accessible in
mean-field theory.  One simply does, or imagines doing, a TDMFT calculation.
Any quantity which one both knows how to compute and to relate to an
experimentally accessible quantity is fair game. To make this concrete we 
should consider the following experimental observable. If we scatter two
nucleons at moderately high energy ($E \sim N_c$), one will generally produce a
final  state with a number of mesons along with two baryons (and at high
enough energies possibly additional baryon-anti-baryon pairs). One can do the 
following experiment---fix a detector at angle, $\theta$ relative to the beam
axis and measure whether an outgoing baryon (or anti-baryon) hits the
detector. By dividing the number of baryons striking the detector minus 
the number of anti-baryons in solid angle $d \Omega$  
per unit time by the incident flux, one obtains a semi-inclusive  
partial differential cross section, $d \sigma^B/d \Omega$.   
Note that this quantity is semi-inclusive; while we know the position of  
one baryon, it does not specify its correlations with the other baryon,  
nor does it specify correlations with outgoing mesons.  As such  
this quantity averages over multiple final states and is not associated  
with any single S-matrix element.  (In fact, with the kinematics restricted 
to $p \sim N_c$, the total amplitude for producing anti-baryons  
is both exponentially suppressed in $N_c$ \cite{LN2} and is experimentally  
small so one can simply use the semi-inclusive baryon differential
cross section in place of the difference of the baryon from the anti-baryon. 
At higher energies one would have to use the difference of the two.) 
 
It is easy to see how to compute this in TDMFT. First consider  
using the Skyrme model at the classical level. While the details of the Skyrme
model need not reproduce QCD, it is thought that the Skyrme model captures all
of the model independent large $N_c$ results from QCD. The model is given in
terms of the dynamics of a chiral  field $U(r,t)\, \in \,SU(2)$ with baryons
treated as solitons. We can fix initial conditions $U(r,0)$ and
$\dot{U}(r,0)$ corresponding to widely separated two hedgehogs with
orientations $A_1$ and $A_2$ moving with a velocity of $\pm v \hat{n}$  
(with $v=\sqrt{2 M_N E}$) and with their centers offset from the $\hat{n}$  
axis by $\pm \vec{b}/2$.  Solving the equations of motion gives
$U(r,t; A_1, A_2, \vec{b}, \hat{n}, E)$ where $A_1, A_2,\vec{b}, \hat{n}, E$
parameterize the initial state. In the Skyrme model the baryon current is
given by the standard topological current, 
\begin{equation}
B_\mu= {\epsilon_{\mu\nu\alpha\beta} \over 24 \pi^2 }
Tr [(U^{-1}\partial^\nu U)(U^{-1}\partial^\alpha U)(U^{-1}\partial^\beta U)] 
\, .
\end{equation}
Using $U(r,t; A_1, A_2,\vec{b}, \hat{n}, E)$ gives the baryon current as a
function of time and space as parameterized by the initial conditions:   
$B_\mu(\vec{r},t; A_1, A_2,\vec{b}, \hat{n}, E)$.  Now consider a large sphere
of radius $R$ centered on the nominal classical collision point of the two
solitons (which we will take to be the origin). The radius $R$  
can be taken to be large compared to all distance scales in the problem. 
The average baryon number per unit solid angle flowing at fixed direction  
(specified by polar angle $\theta$ and azimuthal angle $\phi$) 
out of a single collision is simply given by  
\begin{eqnarray} 
&{}&\frac{d n^B(\theta,\phi)}{d \Omega} \, = \nonumber \\ 
&{}& R^{-2} \int_{- \infty}^{\infty}  dt \, 
\hat{r}(\theta, \phi) \cdot \vec{B}(R \hat{r}(\theta, \phi),t; A_1, A_2, 
\vec{b}, \hat{n}, E)  \, .
\end{eqnarray} 
Integration of $d n^B(\theta,\phi)/d \Omega \,$ 
over $A_1$, $A_2$ and $\vec{b}$ as in Eqs.(\ref{exp}) and (\ref{sig}) 
yields the physical observable 
\begin{eqnarray} 
&{}&\big(\frac{ d \sigma^{B}(\theta,\phi) }{d \Omega}\big )_{m^{(1)}  
{m_I}^{(1)} m^{(2)} {m_I}^{(2)}}  
\, = \nonumber \\  
&{}&  \int {\rm d^2} \vec{b} \, 
\int \, {\rm d}A_1 {\rm d}A_2\, 
\left |D^{1/2}_{m^{(1)},{m_I}^{(1)}}(A_1) \right |^2  
\left | D^{1/2}_{m^{(2)},{m_I}^{(2)}}(A_2) \right |^2 \, \nonumber \\ 
&{}& \times {1\over R^2} \int_{- \infty}^{\infty} \, dt  
\hat{r}(\theta, \phi)  
\cdot \vec{B}(R \hat{r} 
(\theta, \phi),t; A_1, A_2,\vec{b}, 
\hat{n}, E)  
\label{result} 
\end{eqnarray} 
plus  corrections higher order in $N_c$. While the proceeding derivation was
done in the context of the Skyrme model it is clear that we would get the same
structure in any mean-field theory including a direct time-dependent Hartree
calculation (were that technically possible). And thus the form obtained is,
in fact, model-independent. The semi-inclusive differential cross section,
$d \sigma^{B}/d \Omega$, is given by the generalization of Eq.~(\ref{obs1}).
The spin averaged semi-inclusive differential cross section then is given by
$A_0$ and is isospin independent.

From this example it is clear how to construct other observables accessible in
TDMFT. One can construct a variable measuring the fraction of total energy 
flowing out at a fixed angle. However, the most commonly used observables to
describe nucleon-nucleon scattering---the total cross section and the total
and differential elastic cross sections---are not amenable to mean-field
treatments. Elastic cross sections are not calculable since they are directly
associated with particular S-matrix elements. Moreover, it seems impossible to
imagine any way in mean-field theory to extract the elastic cross section. The
total cross section includes forward scattering and, as is well known,
semi-classical treatments fail for forward scattering \cite{Landau}.

Next we consider possible corrections to the spin-isospin structure of  
Eq.~(\ref{obs1}) and its generalizations.  One difficulty with following
Witten's strategy and beginning with TDMFT as giving the leading order result
is that we have not systematically formulated the full $1/N_c$ expansion.
Nevertheless, it seems apparent that while generic $1/N_c$ terms can affect
the values of the various coefficients in Eq.~(\ref{obs1}), they are unlikely
to change the spin-isospin structure. The spin-isospin structure came about
when we neglected terms sensitive to the initial rotational velocities 
of the separated hedgehogs which in turn implied that each explicit power of  
$I$ or $S$ came with a factor of $1/N_c$. If we are considering observables  
associated with operators of good parity and isospin, the spin or  
isospin can only come in as pairs: {\it e.g.} there are no terms proportional 
to $I^{(1)} + I^{(2)}$ but only to $I^{(1)} \cdot I^{(2)}$.  This suggests  
that all corrections to Eq.~(\ref{obs1}) and its generalizations will 
be of order $1/N_c^2$. 
 
It is not clear whether the data on a semi-inclusive differential
cross sections is readily accessible in a form that can be compared to
Eq.~(\ref{obs1}). Presumably such data has been collected at some point.
Comparisons with Eq.~(\ref{obs1}) would be very interesting. It would also be
of interest to see if data on other observables computable from TDMFT can be
found.

\acknowledgements 
 
This work is supported by the U.S.~Department of Energy grant  
DE-FG02-93ER-40762.

\end{document}